\documentclass[12pt,journal,draftcls,letterpaper,onecolumn]{IEEEtran}

\usepackage{cite} 
\usepackage{graphicx}
\usepackage{algpseudocode}
\usepackage{epsfig,epstopdf}
\usepackage{subfigure}
\usepackage{url}
\usepackage{amssymb,amsmath,amsbsy,amsfonts}
\usepackage{expdlist}
\usepackage{color}
\usepackage{rotating}
\usepackage{multirow} 
\usepackage{tabularx}
\usepackage{pdflscape}
\usepackage{booktabs}
\usepackage{scalefnt}

\usepackage{tikz}
\usetikzlibrary{fit,positioning}
\usetikzlibrary{shapes,matrix,decorations.markings,arrows}
\usetikzlibrary{graphs}
\usepackage{pgfplots}

\usetikzlibrary{external} 
\DeclareGraphicsExtensions{.jpg,.eps,.png,.gif,.eps,.ps}
\graphicspath{{./}}

\usepackage{pgfplots}
\usepackage{pgfplotstable}
\usepackage{filecontents}
\usetikzlibrary{plotmarks}

\usepackage{latexsym}
\usepackage{amsmath}
\usepackage{amssymb}
\usepackage{psfrag}
\usepackage{color}
\usepackage{cite}
\usepackage{times}
\usepackage{setspace}
\usepackage{multirow}
\usepackage{url}
\usepackage{hyperref}
\usepackage{bbm}
\usepackage{epstopdf}
\usepackage{graphicx,dblfloatfix}

\def\J{{T}}

\def\rate{R}

\newcommand{\nP}[1]{#1}

\usepackage{mathtools}
\DeclarePairedDelimiter{\ceil}{\lceil}{\rceil}
\DeclarePairedDelimiter\floor{\lfloor}{\rfloor}

\begin{document}

\title{Continuous Transmission of Spatially-Coupled LDPC Code Chains}

\author{\authorblockN{Pablo M. Olmos\thanks{This work was partially supported by NSF grants CCF11-61754 and CCSS-1710920, by TELUS Corporation Canada, by the European Research Council (ERC) through the European Union’s Horizon 2020 research and innovation program under Grant 714161, and by the Spanish Ministry of Economy and Competitiveness and the Spanish National Research Agency under grants TEC2016-78434-C3-3-R (AEI/FEDER, EU) and Juan de la Cierva Fellowship IJCI-2014-19150.}}\thanks{Pablo M. Olmos is with the Departamento de Teor\'ia de la Se\~nal y Comunicaciones, Universidad Carlos III de Madrid. Instituto de Investigaci\'on Sanitaria Gregorio Mara\~n\'on (IiSGM), Madrid, Espa\~na. E-mail: {\tt olmos@tsc.uc3m.es}}, \and \authorblockN{David G. M. Mitchell\thanks{David G. M. Mitchell is with the Klipsch School of Electrical and Computer Engineering, New Mexico State University, Las Cruces, USA.  E-mail: {\tt dgmm@nmsu.edu}}, \IEEEmembership{Senior Member, IEEE},} \\ \and
\authorblockN{Dmitri Truhachev\thanks{Dmitri Truhachev is with the Department of Electrical and Computer Engineering, Dalhousie University, Halifax, Canada.  E-mail: {\tt dmitry@dal.ca}}, \IEEEmembership{Member, IEEE}}, \and \authorblockN{Daniel J. Costello,  Jr.,\thanks{Daniel J. Costello,  Jr. is with the Electrical Engineering Department, University of Notre Dame, Indiana, USA.  E-mail: {\tt costello.2@nd.edu}}
\IEEEmembership{Life Fellow, IEEE}} 
\thanks{This work was presented in part at the  8th International Symposium on Turbo Codes \& Iterative Information Processing (ISTC), August 2014.}}

\usetikzlibrary{external} 
\tikzexternalize[prefix=tikzgraphics/]

\markboth{}{ }
\date{}

\maketitle

\vspace{-1cm}
\begin{abstract}
We propose a novel encoding/transmission scheme called continuous chain (CC) transmission that is able to improve the finite-length performance of a system using  spatially-coupled low-density parity-check (SC-LDPC) codes. In CC transmission, instead of  transmitting a sequence of independent codewords from a terminated SC-LDPC code chain, we connect multiple chains in a layered format, where encoding, transmission, and decoding are performed in a continuous fashion. 
The connections between chains are created at specific points, chosen to improve the  finite-length performance of the code structure under iterative decoding.
We describe the design of CC schemes for different SC-LDPC code ensembles constructed from protographs: a $(J,K)$-regular SC-LDPC code chain,  a spatially-coupled repeat-accumulate (SC-RA) code, and a spatially-coupled accumulate-repeat-jagged-accumulate (SC-ARJA) code. In all cases, significant performance improvements are reported and it is shown that using CC transmission only requires a small increase in  decoding complexity and decoding delay with respect to a system employing a single SC-LDPC  code chain for transmission.
\end{abstract}

\section{Introduction}

Spatially-coupled low-density parity-check (SC-LDPC) codes have attracted a  great deal of interest due to their potential for near-capacity  performance under iterative belief propagation (BP)  decoding \cite{kru13,Kumar14}.  The reason for this behavior is the spatial graph coupling that defines the structure of an SC-LDPC code:  the Tanner graph of a block code with $M$ variable nodes, referred to as the uncoupled LDPC code graph, is replicated $L$ times to produce a sequence of identical graphs; following this, the neighboring copies are then connected to form a chain by redirecting (spreading) certain edges following a chosen coupling pattern. We say that the resulting ``coupled" graph has $L$ positions with $M$ variable nodes each and that it  has a so-called ``structured irregularity'', since parity check nodes located at both ends of the chain are connected to a smaller number of variable nodes than those in the middle \cite{lscz10}. As a result, the nodes at the ends of the graph form strong ``subcodes" and the resulting reliable information generated there during BP decoding propagates through the chain toward the center in a wave-like fashion \cite{kru11}.

The finite-length performance in the waterfall region of a class of  SC-LDPC code ensembles has been analyzed  in \cite{olmos15,olmosmitchell13}, where \emph{scaling laws}, that relate the finite-length block error probability over the binary erasure channel (BEC) to the length of the code and its other structural parameters,  are computed. These results have been extended to SC-LDPC code ensembles generated from protographs in \cite{Stinner15b}, where it is shown that the structure inherited from the the protograph base matrix  improves the finite-length performance. Protograph-based LDPC codes \cite{tho03} possess several practical additional advantages, including smaller decoder memory requirements due to the simplified graph representation, high-speed decoding  utilizing the parallel structure of the graph, and the ability to combine low error floors and good thresholds \cite{ddja09,mitchell14}.

Spatial graph coupling need not be limited to the connection of graphs to form a single chain. In \cite{tru13, CoupChicc2012,CoupChisit2012,tmlc12,Liu15}, more general ensembles were proposed that are constructed by connecting together several individual SC-LDPC code chains. \nP{The resulting structures can  be interpreted as longer SC-LDPC codes with varying coupling patterns.} It was demonstrated that, by optimizing the connection points, the lengths of the connected chains, and their densities, ensembles with improved decoding thresholds can be constructed. \nP{Also, improvements in the iterative decoding convergence speed were observed}. A particularly interesting example  is the loop ensemble~\cite{tru13, CoupChicc2012}, which is constructed by connecting two $(J,K)$-regular SC-LDPC code chains of length $L$ in the form of a loop.
For small to moderate $L$,  loop ensembles have a significantly better BP threshold  than a single component $(J,K)$-regular SC-LDPC code chain of the same rate, while their thresholds coincide as the chain length becomes large. 

From a communication system point of view,  there is growing research interest in developing modulation and transmission techniques that take advantage of the improved performance offered by SC-LDPC coding schemes. For instance, in \cite{Alex15,Aref15} it is shown that optimized bit-mapping of the SC-LDPC coded bits over modulated QAM symbols can be used to enforce stronger termination conditions in the decoding process.
This can be exploited to mitigate the SC-LDPC rate loss, to improve the finite-length performance of the code, or even to improve the BP convergence speed \cite{Aref13}. In this paper, we propose a novel transmission and encoding scheme for SC-LDPC codes that can be regarded as a ``systems'' approach to enhance  finite-length code performance, since our solution affects not only the code design itself, but also the encoding, transmission, and decoding stages.

The proposed method in this paper is  a novel application of SC-LDPC codes based on \nP{terminated} connected chains. In the case of single terminated SC-LDPC  code chains, the entire information squence is split into a number of blocks such that each block is encoded into independent codewords corresponding to unconnected consecutive SC-LDPC chains. We propose to link the different chains and create a continuous stream of encoded information whose Tanner graph is represented by an infinite sequence of connected SC-LDPC code chains. We refer to such an encoding/transmission scheme to as continuous chain (CC) transmission. Note that, we are not connecting chains to form a new block code ensemble, instead we are enforcing a dependence between what before corresponded to independent codewords. We later show that a careful selection of the connection points between chains is critical to boost the CC performance under iterative BP decoding, improving the error rate of single terminated SC-LDPC  code chains.

The underlying principle that explains the performance improvement using CC transmission  is not described using asymptotic (threshold like) arguments, but rather by analyzing the finite-length scaling behavior of a single SC-LDPC code chain and how it is improved when we connect consecutive chains. Explaining this effect is our aim in the first part of the paper, where we show that the finite-length performance of short and long chains is governed by scaling laws with  different characteristics. For long chains, the reliable information generated at the ends of the chain must be propagated towards the center  (the wave-like decoding effect \cite{kru11}), and it can be shown that the block error rate (BLER) scales approximately linearly with the chain length $L$ \cite{olmos15}. However,  for short chains, intermediate positions directly benefit from the reliable information generated at the ends. We show that this  results in an improvement of the finite-length scaling behavior of the code.

In  CC transmission, the key idea is to connect  consecutive SC-LDPC chains  in a way that exploits the two strong subcodes at the ends of each chain to generate reliable information at various points in the graph - effectively breaking the entire connected structure into better protected shorter chains, thereby improving the finite-length performance of the system while maintaining the system coding rate. Compare to the encoding/decoding complexity of independent SC-LDPC chains, CC transmission is feasible for encoding, since it only requires some additional memory in the encoding process,  and also for decoding, where one can effectively implement a windowed decoder \cite{lscz10,iyengar11}, requiring only a different order in the sequence of transmitted bits compared to transmitting unconnected chains. To the best of our knowledge, this is a new concept in the field of SC-LDPC codes. In \cite{olmos14-2}, we presented preliminary results of the CC transmission technique using unstructured randomly-generated $(3,6)$-regular code chains. The present paper extends the construction to general protograph-based ensembles. We present the new construction for the case of representative protograph-based SC-LDPC code ensembles: a $(J,K)$-regular SC-LDPC code ensemble, a spatially-coupled repeat-accumulate (SC-RA) code \cite{johnson13}, and a spatially-coupled accumulate-repeat-jagged-accumulate (SC-ARJA) code \cite{mitchell14}. Simulation results for  the BEC and the binary input additive white Gaussian noise (BIAWGN) channel show that the  finite-length  performance  can be significantly improved with respect to a system transmitting independently encoded SC-LDPC chains. \nP{Due to the continuous nature of CC transmission, the proposed technique can also be considered as a candidate for streaming applications, with the potential to improve  on the performance of  unterminated SC-LDPC codes \cite{Mitchell16, Schmalen14,Smith12}}. Further, CC transmission could be combined with the algebraic methods recently proposed to design SC-LDPC codes with larger girths and improved error floors \cite{Liu16,Zhang16}. Finally, the kind of connected chain structures  proposed here for CC transmission could also  be  used to design SC-LDPC codes with unequal error protection (UEP) constraints \cite{Rahnavard07,Poulliat04}.

The paper is structured as follows. In Section \ref{singlechain} we review the construction of different protograph-based SC-LDPC code chains and  discuss  the analysis of their finite-length performance. In Section \ref{CC} we present the CC transmission scheme and focus on the analysis of CC structures that improve the finite-length performance of a system using  a $(J,K)$-regular SC-LDPC code. In Section \ref{CCcap}, we present  CC structures for the SC-RA and SC-ARJA code ensembles.  In Section \ref{feasibility} the feasibility of CC transmission is analyzed. Finally,  we provide some concluding remarks in Section \ref{future},  including potential research directions on this topic.

\section{Finite-length scaling behavior of a SC-LDPC Single Code Chain}\label{singlechain}

We start by considering a single chain SC-LDPC code ensemble constructed by means of protographs. A protograph \cite{tho03}, or projected graph, is a Tanner graph with a relatively small number of nodes.
It can be also represented in compact form by its bi-adjacency matrix $\mathbf{B}$, called the \emph{base matrix}.
From a protograph with $c$ check nodes and $v$ variable nodes, an $Nc\times Nv$ parity check matrix $\mathbf{H}$ can be derived.
A lifting procedure with lifting factor $N$ replaces each one in $\mathbf{B}$  by an $N\times N$ permutation matrix and each $0$ by an $N\times N$ all-zero matrix.\footnote{Integer entries larger than one in $\mathbf{B}$, representing multiple edges between a pair of nodes, are replaced by a sum of non-overlapping permutation matrices.} An LDPC block code ensemble consists of the set of all possible matrices $\mathbf{H}$ derived from all possible combinations of $N\times N$ permutation matrices.
Since the Tanner graph of $\mathbf{H}$ inherits the degree distribution and graph neighborhood structure of the protograph, the \emph{design rate} of the ensemble can be directly computed from the protograph itself.

\subsection{SC-LDPC protographs}

Several  families of protograph-based SC-LDPC code ensembles have been proposed in the literature, all with different trade-offs between BP thresholds and minimum distance growth rate properties. As  representative members,  we  consider the following three constructions: $(J,K)$-regular SC-LDPC codes \cite{lscz10}, SC-ARJA codes \cite{mitchell14}, and SC-RA codes \cite{johnson13}. Due to its simplicity and well-understood behavior, we first demonstrate the CC transmission scheme for the $(J,K)$-regular SC-LDPC code. We selected the SC-ARJA code ensemble  due to its  asymptotic near-capacity performance, and  we considered the SC-RA code ensemble since it has been reported to possess robust finite-length scaling behavior \cite{Stinner15b}. Our main goal in this paper is  to illustrate the implementation of the CC transmission technique for different \nP{practically relevant} SC-LDPC code ensembles and demonstrate the resulting gains in finite-length performance that can be achieved. As such, the results and designs presented should be regarded as a proof-of-concept of the CC technique.

\begin{figure}[t]
\centering
\begin{tabular}{cc}
\includegraphics[scale=1.0]{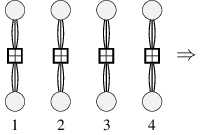}
&
\hspace{-0.25cm}
\includegraphics[scale=1.0]{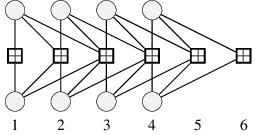}
\end{tabular}
\caption{$\mathcal{C}(3,6,L)$ protograph for $L=4$. The protograph has $L+2$ positions, labelled as $1,2, \ldots, L+2$ from left to right. }\label{figure36}
\end{figure}

The protograph of a $(J,K)$-regular SC-LDPC code can be constructed by coupling  $L$ $(J,K)$-regular block code protographs together, where $J$ is the variable node degree, $K$ is the check node degree, and $L$ is referred to as the chain length \cite{lscz10}. Fig. \ref{figure36} illustrates the construction of  a $(3,6)$-regular SC-LDPC code protograph 
by coupling together a chain of $L=4$ uncoupled $(3,6)$-regular block code protographs. To generate the coupled protograph, edges from variable nodes are spread to  check nodes at neighboring locations, and a ``structured irregularity" is created: the check nodes at the start and the end of the chain are connected to only  2 or 4 variable nodes, while  all the intermediate check nodes have degree six. All  the variable nodes still have degree 3. We denote the SC-LDPC code ensemble generated by lifting this coupled protograph as $\mathcal{C}(3,6,L)$. Finally, note that, due to the termination, there is a \emph{rate loss} compared to the uncoupled $(3,6)$-regular LDPC block code, which has design rate $R=1/2$. Since the protograph in Fig. \ref{figure36} contains $2L$ variable nodes and $L+2$ check nodes, the $\mathcal{C}(3,6,L)$ design rate is $\rate(L)=1-(L+2)/2L=1/2-1/L$.

 \begin{figure}[t]
\centering
\begin{tabular}{cc}
\includegraphics[scale=1.0]{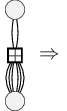}
&
\hspace{-0.5cm}
\includegraphics[scale=1.0]{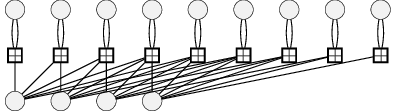}
\end{tabular}
\caption{$\mathcal{C}_{\text{RA}}(q,L)$ protograph for $L=4$ and $q=6$.}\label{figureRA}
\end{figure}

\begin{figure*}[th]
\centering
\begin{tabular}{cc}
\includegraphics[scale=1.0]{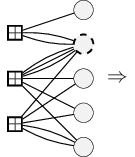}
&
\hspace{-0.5cm}
\includegraphics[scale=1.0]{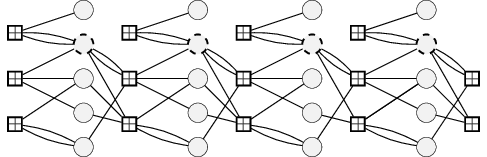}
\end{tabular}
\caption{$\mathcal{C}_{\text{ARJA}}(L)$ protograph for $L=4$.}\label{figureARJA}
\end{figure*}

 In Fig. \ref{figureRA}   we show the protograph of an uncoupled RA code, which contains one degree-two variable node (the accumulator node) and a degree-$q$ variable node (the repetition node). Note that $q=6$ in Fig. \ref{figureRA}. An SC-RA protograph is obtained by coupling $L$ RA protographs by spreading the edges connected to the degree-$q$ variable nodes, as indicated in Fig. \ref{figureRA}. This ensemble is denoted as $\mathcal{C}_{\text{RA}}(q,L)$. Following \cite{johnson13},  $q-1$ accumulator variable nodes are added to the coupled protograph, thus mitigating the rate-loss incurred and avoiding a degree-one check node at the end. This coupling creates, along with lifting, SC-RA code ensembles with near capacity-achieving properties as $q$ increases.  In Fig. \ref{figureARJA}, we show the protograph of  an uncoupled ARJA code and its spatially-coupled counterpart for $L=4$. We denote the coupled code ensemble by $\mathcal{C}_{\text{ARJA}}( L)$. The dashed variable nodes represent \emph{punctured} symbols, i.e.,  associated bits that are not transmitted.

 Table \ref{tab} summarizes the design rates  of each of the three SC-LDPC code ensembles discussed above, as well as their respective iterative decoding  thresholds $\epsilon^*$ over the BEC,  computed for $L=50$. For larger chain lengths, the thresholds are numerically indistinguishable from these values. Note that, in the limit $L\rightarrow\infty$, all the ensembles have rate $1/2$.

\begin{table}[h!]
\caption{Design rate $\rate(L)$ and BP threshold $\epsilon^*$ over the BEC for different SC-LDPC code ensembles}\label{tab}
\begin{center}
\scalebox{1}{%
\begin{tabular}{|lll|}\hline
Ensemble & $\rate(L)$ &  $\epsilon^{*}$ $(L=50)$\\\hline
$\mathcal{C}(3,6,L)$ &  $1-(L+2)/2L$ & $0.4881$\\ 
$\mathcal{C}_{\text{RA}}(4,L)$& $1-(L+3)/(2L+3)$ &  $0.4846$\\
$\mathcal{C}_{\text{RA}}(5,L)$& $1-(L+4)/(2L+4)$& $0.4910$\\
$\mathcal{C}_{\text{RA}}(6,L)$ & $1-(L+5)/(2L+5)$& $0.4934$\\
$\mathcal{C}_{\text{ARJA}}(L)$ & $1-(L+1)/2L$ & $0.4996$\\
\hline
\end{tabular}
}
\end{center}
\end{table}

\subsection{Scaling behavior of a SC-LDPC code chain over the BEC}\label{SLsec}

Analysis of the finite-length performance of  LDPC codes is typically performed over the  BEC \cite{Amraoui2009}. For the BEC, we consider an equivalent formulation to BP decoding called peeling decoding (PD). We start the PD algorithm by removing all the variable nodes and edges associated with non-erased symbols,  plus any disconnected check nodes from the graph. At each iteration,  PD looks for a degree-one check node, which is removed along with the variable node it is connected to. In \cite{lmss01b},  it was shown that if we apply PD to an LDPC code graph, the sequence of graphs produced at consecutive decoding iterations follows a typical path or expected evolution.  In the finite-length regime, the variance around the graph expected  evolution was first derived in \cite{Amraoui2009} for $(J,K)$-regular and Poisson-type LDPC code ensembles. Let $r_1(\ell)$ be the process that represents the evolution along decoding of the fraction of degree-one check nodes in the graph  at iteration $\ell$, and $\hat{r}_{1}(\ell)$ and $\delta_1(\ell)$  its mean and variance, respectively. An estimate of the PD block error probability is obtained by computing the cumulated probability that $r_1(\ell)\leq0$ at those points where $\hat{r}_{1}(\ell)$ presents a local minimum \cite{Amraoui2009}. These points in time are called \emph{critical points}. The resulting estimate to the PD block error probability is called the \emph{scaling law} for the LDPC code ensemble. Following a similar procedure, scaling laws for SC-LDPC code ensembles have been recently derived in \cite{olmos15}  for unstructured randomly-constructed SC-LDPC code chains  and in \cite{Stinner15b} for protograph-based SC-LDPC code chains. 

To describe the phenomena that explains the gain in performance obtained using CC transmission schemes, it is important to review some basic aspects regarding the scaling behavior of  SC-LDPC code chains. To this end,  it suffices to analyze $\hat{r}_{1}(\ell)$, i.e., the expected evolution of the fraction of degree-one check nodes in the graph. We refer to \cite{Stinner15b} for a detailed description of the computation of the function $\hat{r}_{1}(\ell)$ for an arbitrary SC-LDPC protograph. 

The SC-LDPC code block length is $n=v_{\text{unc}}LN+aN$, where $v_{\text{unc}}$ is  the number of variables in the uncoupled protograph, $a=0$ for the ensembles $\mathcal{C}(J,K,L)$ and $\mathcal{C}_{\text{ARJA}}(L)$, but $a=(q-1)$ for the $\mathcal{C}_{\text{RA}}(q,L)$  ensemble. Define the normalized decoding time $\tau=\frac{\ell}{v_{\text{unc}}N}$. Since the average number of erased bits, and hence the average number of iterations, is $n\epsilon$, if follows that
$\tau\in[0,\Omega(L))$, where 
\begin{align}\label{omega}
\Omega(L)=\epsilon L+\epsilon \frac{a}{v_{\text{unc}}}.
\end{align}

\begin{figure}[t]
\begin{center}
\includegraphics[scale=1.0]{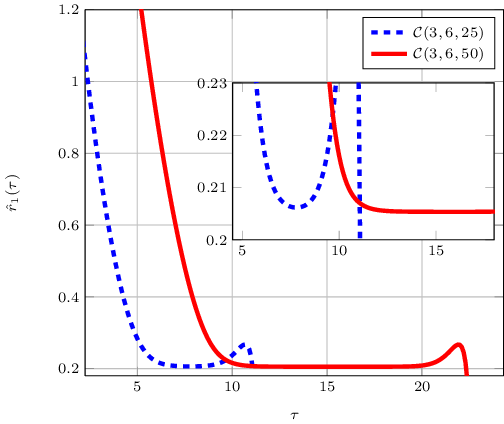}
\end{center}
\caption{Solution to $\hat{r}_{1}(\tau)$ for the ensembles $\mathcal{C}(3,6,25)$ (dashed line) and $\mathcal{C}(3,6,50)$ (solid line) for $\epsilon=0.45$.}\label{Fig: r1chain}
\end{figure}

In Fig.~\ref{Fig: r1chain}, we plot the function $\hat{r}_{1}(\tau)$ for the ensembles $\mathcal{C}(3,6,50)$ (solid line) and $\mathcal{C}(3,6,25)$ (dashed line) and $\epsilon=0.45$. Note that $\hat{r}_1(\tau)$ for  $L=50$  does not display a single critical point, but rather a critical phase  in which  $\hat{r}_1(\tau)$ remains constant; denote this value by $\hat{r}_1(\tau^*)$. During the critical phase, reliable information generated at the boundary positions propagates at constant speed through the chain to the middle positions, and the decoder might fail  at any intermediate point with uniform probability \cite{kru11, Aref15}. In \cite{olmos15}, it is shown that the block error probability during the critical phase can be  estimated as follows:
\begin{align}\label{longchain}
P^*&\approx 1-\exp\left(-\frac{\Omega(L)-\tau^{\circ} }{\displaystyle \frac{\sqrt{2\pi}}{\theta}\int_{0}^{\alpha\sqrt{N}(\epsilon^*-\epsilon)}\Phi(z)\text{e}^{\frac{1}{2}z^2}\text{d}z}\right),
\end{align}
where
\begin{itemize}
\item  $\Omega(L)-\tau^{\circ}=\epsilon L+\epsilon \frac{a}{n_u}-\tau^{\circ}$ is the temporal length of the critical phase, where $\tau^{\circ}$ depends on the particular SC-LDPC ensemble;
\item $\epsilon^*$ is the BP decoding threshold;
\item  $\alpha = \lim_{\epsilon\rightarrow \epsilon^*} \frac{\hat{r}_1(\tau^*)}{\sqrt{\delta_1(\tau^*)}(\epsilon^*-\epsilon)}$, where $\delta_1(\tau^*)$ is the variance of the $r_1(\tau)$ process during the critical phase;
\item $\Phi(z)$ is the  c.d.f. of the standard Gaussian distribution, $\mathcal{N}(0,1)$; and
\item $\theta$ is  a parameter related to the exponential decay of the covariance of  $r_1(\tau)$ with time, i.e., $\text{COV}[r_1(\tau), r_1(\zeta)]\propto \exp(-\theta |\zeta-\tau|)$.
\end{itemize}
The scaling parameters $\alpha$ and $\theta$  depend on the degree distribution and coupling pattern of the SC-LDPC code ensemble,  but they are independent of the chain length $L$. (The actual computation of the different parameters in \eqref{longchain} is not relevant  here, see \cite{olmos15} for further details.) An important result that can be derived from \eqref{longchain} is that, for large $N(\epsilon^*-\epsilon)$, 
\begin{align}\label{longchain_large}
P^*&\approx \frac{1}{\sqrt{2\pi}}\frac{\alpha \theta \epsilon L}{\sqrt{N}(\epsilon^*-\epsilon)}\mathrm{e}^{\displaystyle -\frac{N(\epsilon^*-\epsilon)^2}{\alpha^2}}\quad \text{ as } N(\epsilon^*-\epsilon)\rightarrow \infty,
\end{align}
and hence in the low error rate regime the SC-LDPC code BLER scales roughly linearly with the chain length $L$. This result is expected due to the convolutional structure of the code.

  \begin{figure}[t!]
\begin{center}
\includegraphics[scale=1.0]{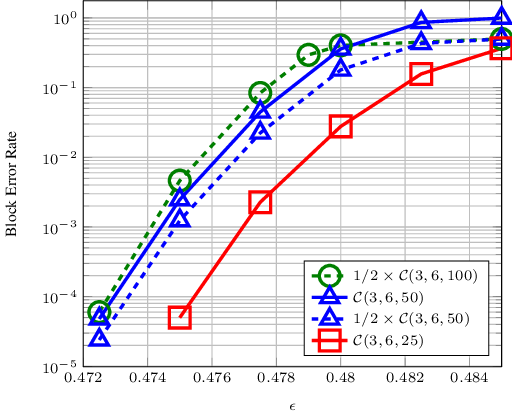}
\end{center}
\caption{Solid lines represent the simulated BLER for $N=1000$ of the ensembles $\mathcal{C}(3,6,L)$ for $L=25$ $(\square)$ and $L=50$ $(\triangle)$, while dashed lines represent $0.5$ times the simulated BLER for the $\mathcal{C}(3,6,L)$ ensemble with $L=50$ and $L=100$. }\label{Fig: scaling}
\end{figure} 

\begin{figure}[t!]
\begin{center}
\begin{tabular}{c}
\includegraphics[scale=1.0]{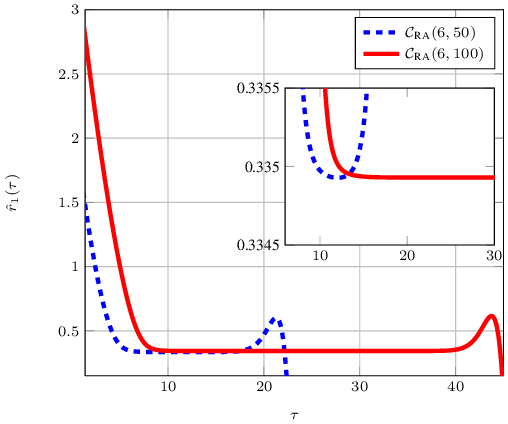} \\ (a) \\ 
\includegraphics[scale=1.0]{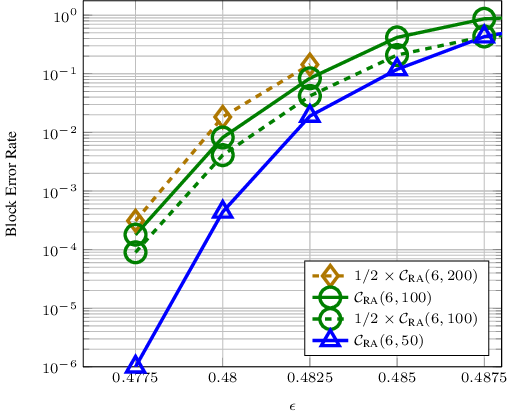}\\
(b)
\end{tabular}
\end{center}
\caption{In (a), we plot  to $\hat{r}_{1}(\tau)$ for the ensembles $\mathcal{C}_{RA}(6,50)$ (dashed line) and $\mathcal{C}_{RA}(6,100)$  (solid line) for $\epsilon=0.45$. In (b), we represent  the simulated BLER for $N=1000$ of the ensembles $\mathcal{C}_{RA}(6,L)$ for  $L=50$ $(\triangle)$ and $L=100$ $(\circ)$ using solid lines. With dashed lines we represent $0.5$ times the error rate computed for the $\mathcal{C}_{RA}(6,L)$ ensemble  for $L=100$ and $L=200$. }\label{FigsRA}
\end{figure}

The scaling behavior model in \eqref{longchain} is only valid as long as the decoding process is governed by the wave-like propagation of reliability. After magnification, we  observe in  Fig. \ref{Fig: r1chain}  that $\hat{r}_{1}(\tau)$ for the $\mathcal{C}(3,6,25)$ ensemble presents a single critical point at $\tau^*\approx7.4$ rather than a critical phase. Such a single critical point corresponds to that point in time when the graph is in the following state: the central positions in the graph have no degree-one check nodes, which can be found only at the ends of the graph; however, for small chain lengths $L$, as soon as the variable nodes connected to the low-degree check nodes at the ends are successfully decoded, a large fraction of degree-one check nodes is created along the entire  chain and decoding succeeds with very high probability. In other words, there is no decoding wave traveling along the chain. In this case, the variable nodes at the central positions in the chain directly benefit from the effect of the two stronger subcodes at both ends, and this increases the robustness of the decoding process beyond what is predicted by \eqref{longchain}.  
In this case, in \cite{olmosmitchell13,olmos14-2} it  was shown that the scaling law that explains the relation between block error probability and the parameters of the code is given by
\begin{align}\label{Q}
P^*&\approx \mathcal{Q}\left(\frac{\sqrt{N}(\epsilon^*-\epsilon)}{\alpha}\right),
\end{align}
where $\mathcal{Q}\left(\cdot\right)$ is the complementary error function of Gaussian statistics. Note that, for fixed $\alpha$, \eqref{Q} predicts  better scaling between BLER and lifting factor $N$ than \eqref{longchain_large}, since it does not depend directly on the chain length $L$.\footnote{The values computed for $\alpha$ in both scenarios are not equal, but minor differences were observed in \cite{olmosmitchell13}.} 

For instance, consider  the ensemble  $\mathcal{C}(3,6,L)$ for $L=25,50$, and $100$ with $N=1000$. The BP thresholds computed for these three chain lengths are identical up to the 5th decimal place, $\epsilon^*\approx 0.48815$.  According to \eqref{longchain_large},  for $\epsilon$ far from the threshold ($\epsilon\ll\epsilon^*$) we should observe a linear degradation of the block error rate as we increase the chain length $L$.  In Fig. \ref{Fig: scaling} we represent the simulated block error rates for these three ensembles using solid lines. With dashed lines, we represent  the BLER computed for  $L=50$ and $L=100$, multiplied by a factor of $1/2$, which gives the estimated performance of the ensembles $\mathcal{C}(3,6,25)$ and $\mathcal{C}(3,6,50)$, respectively, according to \eqref{longchain_large}. Note that, while the estimated performance for the $L=50$ case is reasonably accurate, in the $L=25$ case the simulated BLER is substantially better than the one  predicted using \eqref{longchain_large}, since the  decoding scaling behavior is ruled by a different and more robust behavior function in this case.

Experiments carried out  for different SC-LDPC code ensembles indicate the same effect is  always  observed, even though the transition between the finite-lengh models (single critical point versus critical phase) can occur at different values of $L$, depending on the code ensemble. For instance, in Fig. \ref{FigsRA}(a) we show the $\hat{r}_{1}(\tau)$ solution for the $\mathcal{C}_{RA}(6,L)$ ensemble with $L=50$ and $L=100$ at $\epsilon=0.45$, where a single critical point can be observed in the $L=50$ case.  Simulation results for $N=1000$  are included in Fig. \ref{FigsRA}(b), and they confirm that the BLER achieved by the $L=50$ chain is substantially better than the BLER simulated for the $L=100$ chain and corrected by  a  factor $1/2$. Note again that the estimate is accurate, however, if we use  the BLER computed for $L=200$ to estimate the performance for $L=100$.

\section{Continuous-chain transmission for \textcolor{black}{$(3,6)$-regular SC-LDPC code chains}}\label{CC}

Design rate is typically a fixed constraint in many communication systems, and thus the use of  short (lower rate) SC-LDPC code chains that exhibit more favorable finite-length properties may not be a viable option. Thus, $L$ is usually chosen large enough to mitigate the rate loss due to the termination. As an alternative, we present a novel  transmission scheme, referred to as continuous chain (CC) transmission  of SC-LDPC codes, that improves the finite-length performance of a system using a long SC-LDPC code chain with minimal rate loss, i.e., a system where the finite-length performance would normally be predicted via the scaling law in \eqref{longchain}. CC transmission can be  regarded as a systems-oriented solution in the sense that its deployment requires changes in both the encoding, transmission, and decoding stages.  In this section, we present the fundamentals of CC transmission and illustrate it for a system using a $\mathcal{C}(J,K,L)$ code chain with $J=3$ and $K=6$. In Section \ref{CCcap}, we  extend the discussion to the $\mathcal{C}_{\text{RA}}(q,L)$ and  $\mathcal{C}_{\text{ARJA}}(L)$ code chains. \nP{The designs of the CC structures discussed in these two sections have been obtained after extensive heuristic optimization, with the aim of maximizing the robustness at those points where we connect different SC-LDPC code chains. This latter aspect is formalized by the notion of a \emph{region threshold}, introduced in Section \ref{regth}.}

\subsection{The CC structure and protograph}

\begin{figure}[t]
\includegraphics[scale=1.25]{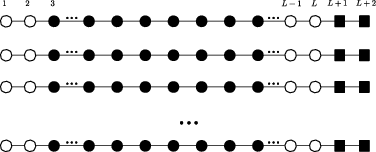}
\caption{ \nP{Schematic representation} of  independent $\mathcal{C}(3,6,L)$ code chains.}\label{ind36}
\end{figure}

Consider first the encoding and transmission of codewords using a  single $\mathcal{C}(3,6,L)$ code chain as an example.  The information stream is divided into blocks of $n\rate(L)$ bits, where the design rate $\rate(L)$ is given in Table \ref{tab}. These blocks are then independently encoded, transmitted, and decoded at the receiver. In the following, by transmitting independent code chains we mean the transmission of independent  codewords that belong to a particular member of the ensemble. Fig. \ref{ind36} shows a \nP{schematic representation} of independent $\mathcal{C}(3,6,L)$ code chains, where the round dots indicate positions with variable nodes, white dotes indicate positions where the variable nodes are connected to check nodes of degree less than $6$, i.e., they represent regions of better protection as a result of lower-degree check nodes, and square dots indicate positions with no variable nodes. 

The main idea behind CC transmission is that the finite-length performance of a system transmitting independent (long) code chains can be  improved if we create a dependency between these chains, where data is encoded in a continuous fashion using a convolutional-like structure based on connected SC-LDPC code chains. This structure, \nP{which  can also be regarded as a single SC-LDPC code chain with a non-standard coupling pattern}, is  hereafter referred to as the \emph{CC structure}. Rather than using independent chains, as shown in Fig.  \ref{ind36}, we connect the chains in such a way  that creates some variable nodes with degree greater than 3, which results in robust   regions of better protection in the middle of each chain that can be succesfully decoded with much higher probability than the remaining positions. In this way, once the robust regions are decoded, each of the remaining and still undecoded chain segments  are short enough to present a finite-length scaling behavior determined by a single critical point, which improves the overall finite-length performance. 

\begin{figure}[t]
\begin{center}
\begin{tabular}{c}
\includegraphics[scale=1.25]{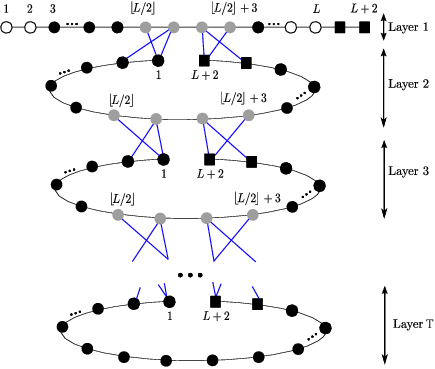}
\end{tabular}
\caption{\nP{Schematic representation} of a CC structure based on connected $\mathcal{C}(3,6,L)$ chains. Gray dots indicate positions with variable nodes of degree greater than $3$.}\label{CC1}
\end{center}
\end{figure}

 This is precisely the effect achieved by the CC structure illustrated in Fig. \ref{CC1}, in which we say there is a single $\mathcal{C}(3,6,L)$ chain per layer.  Each chain, which in Fig. \ref{ind36} corresponds to an independent codeword, is now jointly encoded with the chain in the layer above and with the chain in the layer below.  Layers of the CC structure are labeled from $1$ to $\J$, and the layer $j$ chain is simply referred to as chain $j$, $j=1,2,\ldots,\J$.  The chains are connected such that the low-degree check nodes at both ends of  chain $j$ in Fig. \ref{ind36} are used to increase the degrees of variable nodes in intermediate positions of chain $(j-1)$, $j=2,3, \ldots,\J$. Gray dots in Fig. \ref{CC1} indicate positions with variable nodes of degree greater than $3$, i.e., variable nodes that are better protected than those in positions represented by black dots. Note that, except for chain $1$, no chain has low-degree check nodes anymore. Also note that chain $\J$ does not  contain any variable nodes with  degree greater than 3 or any low degree check nodes. The connection points are designed such that the variable nodes at positions $\floor{L/2}$ and $\floor{L/2}+3$ of chain $j$ are connected with an extra edge to the check nodes at positions $1$ and $L+2$  of chain $j+1$, respectively, so that  they have degree $4$. Similarly, the variable nodes at positions $\floor{L/2}+1$ and $\floor{L/2}+2$ are connected with two extra edges to the check nodes in the layer below, and hence they have degree $5$. 

Note that creating the CC structure does not change the design rate; rather, we simply exploit the termination rate loss in a different manner.  \nP{For example, CC structure in Fig. \ref{CC1} contains exactly the same number of coded bits as if we transmitted $\J$ consecutive  independent $\mathcal{C}(3,6,L)$ codewords.} Our goal is to demonstrate that the finite-length performance measured per chain layer in Fig. \ref{CC1} is significantly better than the performance per codeword in Fig. \ref{ind36}. 

\begin{figure}[t]
\centering\includegraphics[scale=1]{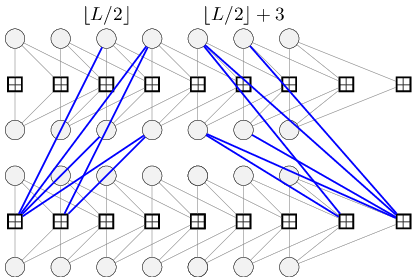}
\caption{CC protograph with $\J=2$ layers using $\mathcal{C}(3,6,L)$ chains with length $L=7$. }\label{CC2}
\end{figure}

The parity-check matrix of a code associated with the CC structure in Fig. \ref{CC1} can be generated by applying the lifting procedure to the CC protograph, which is obtained by connecting $\J$ copies of the $\mathcal{C}(3,6,L)$ protograph in Fig. \ref{figure36} according to Fig. \ref{CC1}.  For instance,  the CC protograph for $\J=2$ and $L=7$ is shown in Fig. \ref{CC2}, where the additional edges placed to connect chains at  layers 1 and 2 are plotted with thick  lines.  Note that, in the upper protograph, the variable nodes at positions $\floor{L/2}$ and $\floor{L/2}+3$ have degree $4$  and those at positions $\floor{L/2}+1$ and $\floor{L/2}+2$ have degree 5, while in the lower protograph, all the variable nodes have degree 3.

\subsection{Asymptotic analysis}\label{regth}

Based on the CC protograph,  we analyze the asymptotic ($N\rightarrow\infty$) performance of the CC structure by computing the expected evolution  of the fraction of degree-one check nodes in the graph under PD \cite{Stinner15b}. The first step in the CC design process is to quantify the robustness of the region created in the middle of each chain, where variable nodes have higher degrees. Recall that our goal is to ensure that this region is succesfully decoded with very high probability, compared to the decoding threshold of the whole structure.  Consider the region between positions  $\floor{L/2}$ and $\floor{L/2}+3$ in chains $1$ to $\J-1$. We define the region  threshold $\epsilon^*_{\floor{L/2},\floor{L/2}+3}$ as the maximum $\epsilon$ value for which all variable nodes at positions $\floor{L/2},\floor{L/2}+3$ in chains $1$ to $\J-1$ are successfully decoded in the  limit $N\rightarrow\infty$.\footnote{The evaluation of the expected graph evolution requires the numerical integration of a very large system of differential equations, and this must be repeated for each $\epsilon$ value, thus becoming a cumbersome and time-consuming task. As a consequence, we adopt the following criterion to numerically estimate the different thresholds: a position in an SC-LDPC code chain is considered decoded if the  fraction of variable nodes undecoded is below $\delta=10^{-3}$.} In addition, we define the CC threshold $\epsilon^*_{\text{CC}}$ as the maximum $\epsilon$ value such that all variable nodes in chains $1$ to $\J-1$ are successfully decoded.

 Table \ref{tabCC} shows $\epsilon^*_{\text{CC}}$ and $\epsilon^*_{\floor{L/2},\floor{L/2}+3}$  for different values of $L$ and the CC structure based on the connected $\mathcal{C}(3,6,L)$ chains  in Fig. \ref{CC1}.  Observe that, as $L$ grows, 
$\epsilon^*_{\text{CC}}$ converges to the threshold of the $\mathcal{C}(3,6,L)$ code ensemble, implying that asymptotically all chains behave similarly. However, note that the region threshold $\epsilon^*_{\floor{L/2},\floor{L/2}+3}$ saturates at $0.502$, above the CC threshold for large $L$. Since $\epsilon\leq\epsilon^*_{\text{CC}}$ is the natural operating region, this implies that positions $(\floor{L/2},\floor{L/2}+1,\floor{L/2}+3)$ in chains $j=1,2,\ldots,\J-1$ are almost surely decoded. Consequently,  connected chains are essentially broken into   chains of length roughly $\floor{L/2}$ in the decoding process.  This means that decoding failures in chains $j=1,2,\ldots,\J-1$, will not propagate to lower chains.  More importantly, for intermediate values of the chain length $L$, the decoding of each of the two segments of length $\floor{L/2}$ is described by the single-critical point model in \eqref{Q}, so we benefit from the much  better  finite-length scaling  properties of these shorter segments.

\begin{table}[t]
\caption{CC threshold $\epsilon^*_{\text{CC}}$ and region  threshold $\epsilon^*_{\floor{L/2},\floor{L/2}+3}$ for the CC structure in Fig. \ref{CC1}.}\label{tabCC}
\begin{center}
\scalebox{1}{%
\begin{tabular}{|ccc|}\hline
$L$ & $\epsilon^*_{\text{CC}}$  &  $\epsilon^*_{\floor{L/2},\floor{L/2}+3}$\\\hline
$10$ &$0.541$ & $0.541$\\
$20$& $0.5015$ &  $0.502$\\
$30$& $0.488$& $0.502$\\
$50$ &  $0.488$& $0.502$\\
\hline
\end{tabular}
}
\end{center}
\end{table}

 \begin{figure}[t]
 \begin{center}
\includegraphics[scale=1]{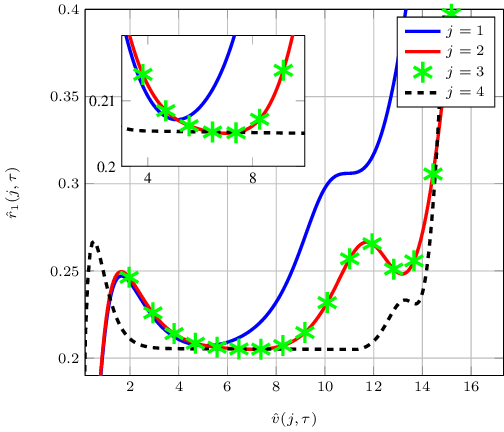}
\end{center}
\caption{Evolution of the normalized number of degree-one check nodes per layer $\hat{r}_{1}(j,\tau)$, $j=1,2,3,4$, as a function of the  normalized number of variable nodes per layer $\hat{v}(j,\tau)$ during PD for  $\epsilon=0.45$ and the  CC structure of Fig. \ref{CC1} and $L=50$.}
\label{Fig_N_4}
\end{figure}

\begin{figure*}[th]
\begin{center}
\includegraphics[scale=1.25]{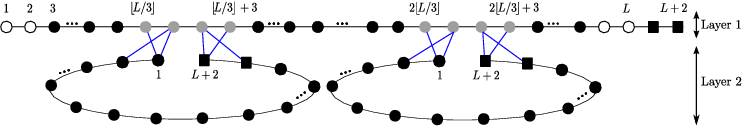}
\caption{\nP{Schematic representation} of a two-layer CC transmission scheme where each  $\mathcal{C}(3,6,L)$ chain is connected to two chains in the layer below.}\label{CC2two}
\end{center}
\end{figure*}

To illustrate this, in Fig. \ref{Fig_N_4}  we plot the expected graph evolution for $\epsilon=0.45$ of the normalized number of degree-one check nodes in each chain $\hat{r}_{1}(j,\tau)$, $j=1,2,3,4$, as a function of the normalized number of variable nodes per chain $\hat{v}(j,\tau)$ for a CC structure with $\J=4$ layers based on $\mathcal{C}(3,6,L)$ chains with $L=50$.  As we  observe from the magnification in the  upper left corner, in contrast to chain $4$, the finite-length performance of chains $1$ to $3$ is determined by a single critical point, suggesting  that the finite-length performance of chains $1$ to $3$ will be significantly improved compared to chain $4$.

\nP{Alternative designs for the connection points yield different values of the region threshold. For instance, if we modify the connecting edges in the CC protograph in Fig. \ref{CC2} to create six consecutive positions in the upper layer (from $\floor{L/2}-1 $ to $\floor{L/2}+4$) with degree 4 variable nodes, the region threshold for $L=50$ drops to $\epsilon^*_{\floor{L/2}-1,\floor{L/2}+4}\approx 0.494$. On the other hand, if only two consecutive positions with variable nodes of degree 6 are created, the region threshold numerically coincides with that of the CC protograph in Fig. \ref{CC2}, i.e., $\epsilon^*_{\floor{L/2},\floor{L/2}+1}\approx 0.502$.}
\vspace{2mm}

When the rate constraints are even stricter and, as a consequence, we must use even longer chains, e.g., $L=100$, the CC structure in Fig. \ref{CC1} might not be able to provide the same gain per layer as for a shorter chain, e.g., $L=50$. A possible alternative is to break each chain of length $L=100$ into three shorter segments  by creating two intermediate regions with stronger protection. We illustrate this construction in Fig. \ref{CC2two} for the case $\J=2$.
Following this approach, note that if we add another layer to the CC structure, we will need four chains to improve the performance of the  chains in layers 1 and  2. For an arbitrary number $\J$ of layers, the ratio of the number of strongly protected chains to the total number of chains is
\begin{align}
\eta=\frac{\displaystyle\sum_{j=1}^{\J-1}2^{j-1}}{\displaystyle\sum_{j=1}^{\J}2^{j-1}},
\end{align}
which  tends  to $1/2$ with increasing $\J$. This implies that, when using CC transmission, half of the chains will enjoy  better performance with no significant increase in  encoding/decoding complexity, as we explain in Section \ref{feasibility}. 

We remark here  that many variations on the CC structure are possible. In fact, alternative structures could be proposed  where the main design goal is to provide unequal error protection for different chains. For instance, we could use the CC structure in Figs. \ref{CC1} and \ref{CC2two}  with only two layers, $\J=2$, to provide additional protection to half of the chains in the first case or to one-third of the chains in the second case.

\subsection{Computer simulation results}

\begin{figure}[t!]
\begin{center}
\includegraphics[scale=1]{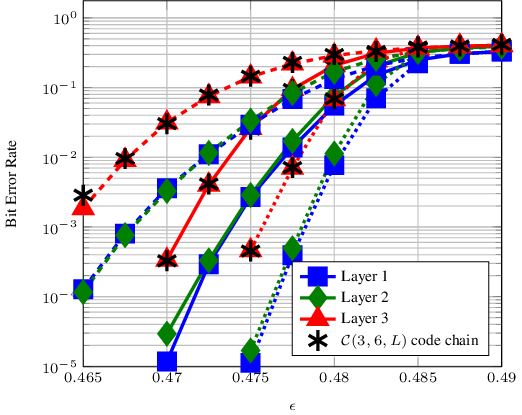}
\end{center}
\caption{BER for the CC structure in Fig. \ref{CC1} with $\J=3$ layers based on $\mathcal{C}(3,6,L)$ chains with $L=50$. The lifting factors are $N=500$ (dashed lines), $N=1000$ (solid lines), and $N=2000$ bits (dotted lines).}\label{figBER36}
\end{figure}

The above conclusions are corroborated by computer simulation. Since we focus on intermediate chain length values $L$, the performance gains achieved by CC transmission can be illustrated in terms of either  BLER or bit error rate (BER) figures. In Fig. \ref{figBER36} we show the BER  per chain for the $\mathcal{C}(3,6,L)$ CC structure in Fig. \ref{CC1} with $\J=3$, $L=50$ and different lifting factors $N$.  We also include the BER measured for a single $\mathcal{C}(3,6,L)$ code chain. As predicted, we obtain a significant gain in performance in chains  1 and 2, of almost one order of magnitude, even for large codes with $4000$ bits per position. Note that there is no degradation in the performance of chain 2 with respect to chain 1,  even though chain 2 does not have low-degree check nodes at each end. This is critical to the success of the CC transmission approach for an arbitrary number of layers, and this behavior has further been confirmed by simulation for ensembles with a larger number of layers. Also, we note  that there is no performance degradation in  chain 3 with respect to a single $\mathcal{C}(3,6,L)$ code chain. This demonstrates the UEP property of CC transmission.

\begin{figure}[h]
\centering\includegraphics[scale=1]{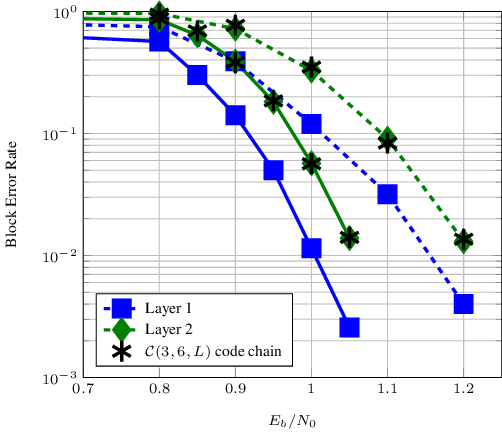}
\caption{BLER for transmission over a BIAWGN channel for the CC structure in Fig. \ref{CC1} with $\J=2$ layers based on $\mathcal{C}(3,6,L)$ code chains with $L=50$. The lifting factors used are $N=250$ (dashed lines) and  $N=500$ (solid lines).}\label{figBER36AWGN}
\end{figure}

SC-LDPC codes have been shown to be \emph{universal}, in the sense that they display capacity-approaching performance across binary-input memoryless channels \cite{kru13, Kumar14}. In  Fig. \ref{figBER36AWGN}, we show exemplary results for the CC structure  depicted in Fig. \ref{CC1} with parameters $\J=2$ and $L=50$  used for transmission over the BIAWGN channel. Note that we observe the same behavior that we did for the BEC. We again observe that the performance of  chain 1  is substantially improved with respect to chain 2, which in turn has the same performance as the single chain of the same length and rate.

\section{CC transmission for SC-RA and SC-ARJA code chains}\label{CCcap}

In this section, we describe the implementation of CC transmission for the $\mathcal{C}_{\text{RA}}(q,L)$ and $\mathcal{C}_{\text{ARJA}}(L)$ ensembles. As described previously, a careful design of the connection points between consecutive chains is crucial for CC transmission, since we need  such points to be characterized by region thresholds above the threshold of the whole CC structure. 

\subsection{Creating robust connection points}

For the $\mathcal{C}_{\text{RA}}(q,L)$ and $\mathcal{C}_{\text{ARJA}}(L)$ ensembles,  both of which contain check nodes of high degree, we have found that to create a region threshold sufficiently above the iterative decoding threshold of each code chain,  we must greatly increase the degree of the variable nodes while maintaining the check node degree distribution. The same issue also arises for the $\mathcal{C}(4,8,L)$ and $\mathcal{C}(5,10,L)$ ensembles and, in general, for capacity approaching SC-LDPC code ensembles with high-degree check nodes. This complicates the design of the connection points, since typically a single code chain per layer will not be able to provide the  necessary number of additional edges to create a robust connection  in the middle of the code chain in the above layer. 

For instance, consider the $\mathcal{C}_{\text{RA}}(6,L)$ protograph shown in Fig. \ref{figureRA}. There are 30 edges that can be connected to the check nodes at the boundary positions without creating a check node of degree higher than $8$. By following a  CC structure similar to that in Fig. \ref{CC1}, we can use  these additional $30$ edges to increase the degree of the repetition variable nodes from $6$ to $12$ in five consecutive positions in each chain. This results in a  region threshold is $\epsilon^*_{\floor{L/2},\floor{L/2}+4}=0.463$, which however is  worse than the BP threshold of the $\mathcal{C}_{\text{RA}}(6,L)$ chain, 
$\epsilon^*=0.4934$.

\begin{figure}[t]
\begin{center}
\includegraphics[scale=1.25]{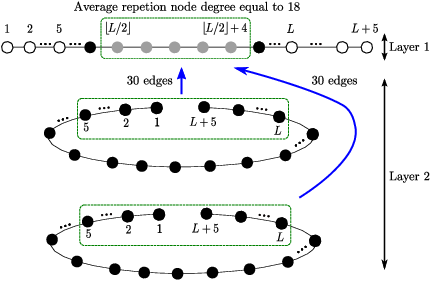}
\caption{\nP{Schematic representation} of a CC structure based on connected $\mathcal{C}_{\text{RA}}(6,L)$  chains. The two chains in layer $2$ are used to create a robust intermediate region in the above chain.}\label{CCRA}
\end{center}
\end{figure}

A possible method to overcome this problem is to use more than one chain per layer, as  illustrated in Fig. \ref{CCRA} for $\J=2$, which uses two chains per layer to protect one chain in the layer above. Two $\mathcal{C}_{\text{RA}}(6,L)$ code chains per layer can provide up to  60 additional edges. If we increase the degrees of the repetition variable nodes from $6$ to $18$  in five consecutive positions (maintaining the degree $2$ accumulator variable nodes), we compute a region threshold $\epsilon^*_{\floor{L/2},\floor{L/2}+4}=0.493$, again not quite matching the $\mathcal{C}_{\text{RA}}(6,L)$ threshold. Following this approach, a third chain per layer would be necessary to further improve the region threshold above this value. As previously discussed, however, two chains per layer reduces the ratio of the number of strongly protected chains to the total number of chains, which for the CC structure in  Fig. \ref{CCRA}  tends  to $\nu=1/2$ as $\J$ increases. If a third chain per layer is used,  this would reduce the ratio to $\nu=1/3$ as $\J$ increases.

\begin{figure}[t]
\includegraphics[scale=1]{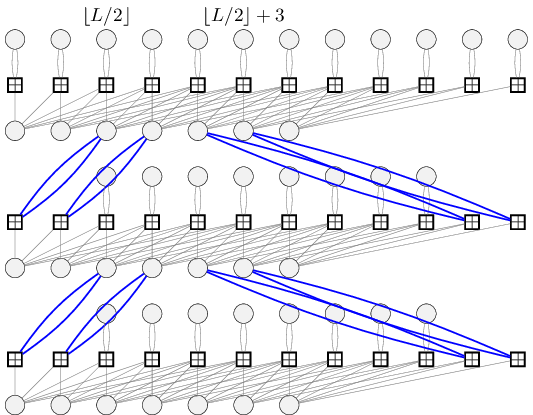}
\caption{CC protograph with a single chain per layer and $\J=3$ layers based on a modified $\mathcal{C}_{RA}(6,L)$ code chain with $L=7$.}\label{CCRAlow}
\end{figure}

Alternatively, a different approach is  to design the CC structure using a modified $\mathcal{C}_{\text{RA}}(6,L)$ code chain with a slightly larger rate loss and, consequently, a code chain that possesses stronger end terminations. An example is shown for the $\mathcal{C}_{\text{RA}}(6,L)$ ensemble in Fig. \ref{CCRAlow}, where we show the protograph of a CC structure with $\J=3$ layers and a single chain per layer. Note that the accumulator variable nodes in two consecutive positions at both ends have been removed and that the corresponding edges are reconnected to repetition variable nodes in the above layer, thereby decreasing the code rate of the chains in the lower layers. All chains have length $L=7$, but the chains in the second and third layers, and in any possible  lower layers, have slightly lower design rate than the $\mathcal{C}_{\text{RA}}(6,L)$ code chain. 
In this CC structure, robust regions in the middle of each layer are created using only $8$ edges  from the chain in the layer below, resulting in a region threshold $\epsilon^*_{\floor{L/2},\floor{L/2}+3}=0.498$, larger than the threshold $\epsilon^*$ of the modified $\mathcal{C}_{RA}(6,L)$ ensemble, which is roughly equal to that of the $\mathcal{C}_{RA}(6,L)$ ensemble, i.e., $\epsilon^*\approx 0.4934$.

\begin{figure}[t]
\centering\includegraphics[scale=1]{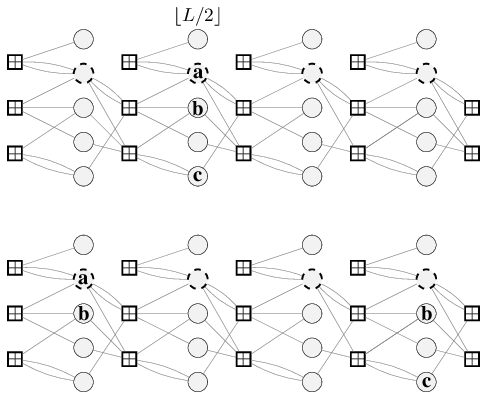}
\caption{CC protograph with a single chain per layer and $\J=2$ layers based on a modified  $\mathcal{C}_{\text{ARJA}}(L)$ code chain with $L=4$.}\label{CCARJA2}
\end{figure}

Designing a CC structure for the $\mathcal{C}_{\text{ARJA}}(L)$ code ensemble follows a similar procedure, and robust intermediate connections can be achieved either by using more than one chain per layer to obtain the necessary additional edges or by slightly decreasing the rate of the code chain. Indeed, for the $\mathcal{C}_{\text{ARJA}}(L)$ code ensemble, a moderate increase in the rate loss is less of a factor since it possesses the  smallest rate loss compared to the other SC-LDPC code ensembles considered (see Table \ref{tab}). In Fig. \ref{CCARJA2}, we show the CC protograph for $\J=2$ layers with a modified lower-rate $\mathcal{C}_{\text{ARJA}}(L)$ code chain in each layer. In the figure, to improve readability, variable nodes with the same letter represent the same variable node, whose degree is the sum of the individual degrees. For example, node $a$ has degree 12, node $b$ has degree $9$, and node $c$ has degree $6$. Recall also that dashed variable nodes indicate punctured symbols.

For all the cases discussed above, note that the design of the connection points is not unique and may be suboptimal. In fact, there might be different configurations that give rise to similar region thresholds.  In general, to design  CC structures like those proposed above, we aim to increase the degrees of the variable nodes in a certain region of a chain by exploiting low-degree check nodes at the end terminations of subsequent code chains. We validated our designs using by combining the CC protograph with the asymptotic analysis in \cite{Stinner15b} to confirm that the resulting local threshold was above the threshold of the SC-LDPC code chain. As long as this condition is met, we did not find a significant difference  in performance between a robust region that spans for several positions or a shorter robust region with very high-degree variable nodes. For instance, in both 
the CC structure based on the $\mathcal{C}(3,6,L)$ ensemble in Fig. 8 and the CC structure based on the $\mathcal{C}_{\text{RA}}(6,L)$  ensemble in Fig. 15, we  observe a robust region of four consecutive positions with variable nodes of higher degree. On the other hand, for the CC structure of the $\mathcal{C}_{\text{ARJA}}(L)$ ensemble in Fig. 16, the robust region is formed by a single position.

Also, we found that increasing the check node degrees at the chain terminations above the degrees of the original uncoupled LDPC block code did not result in viable configurations, since the region threshold at the connection points was significantly worse that the SC-LDPC code chain decoding threshold. As explained above, if additional edges are needed to create robust regions, we found that using several chains and/or increasing the rate-loss at chain terminations are more effective approaches.

\subsection{Computer simulation results}

\begin{figure}[ht!]
\begin{center}
\begin{tabular}{c}
\includegraphics[scale=1]{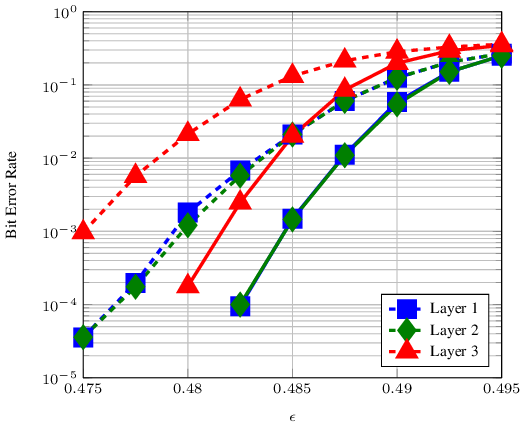} \\
(a) \\
\includegraphics[scale=1]{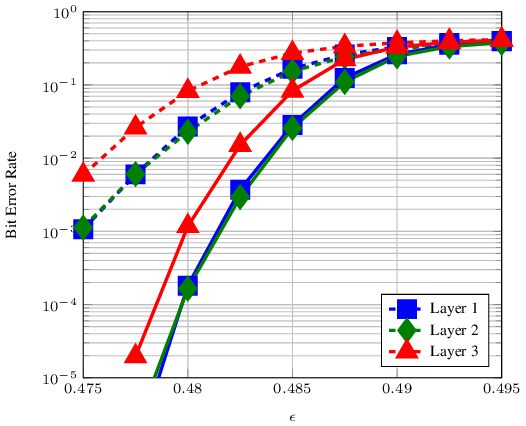} \\
(b)
\end{tabular}
\caption{BER performance for a CC structure  with $\J=3$ layers based on the modified $\mathcal{C}_{RA}(6,L)$ code chain  with (a) $L=50$ and (b) $L=100$. The lifting factors are $N=500$ (dashed lines) and $N=1000$ (solid lines). }\label{BERRA}
\end{center}
\end{figure}

\begin{figure}[h]
\centering
\centering\includegraphics[scale=1]{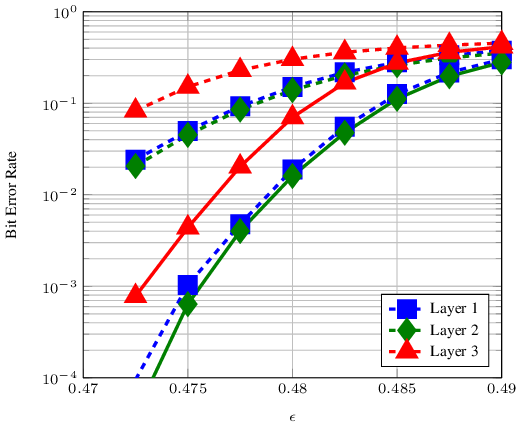}
\caption{BER performance for a CC with $\J=3$ layers based on a modified $\mathcal{C}_{\text{ARJA}}(L)$ code chain with $L=50$. The lifting factors are $N=250$ (dashed lines) and $N=500$ (solid lines).}\label{BERARJA}
\end{figure}

Once we have created robust regions in the intermediate positions of each chain, the phenomenon that explains the CC gain in performance occurs as described in Section \ref{CC}. Fig. \ref{BERRA} shows the simulated BER performance for a CC structure  with $\J=3$ layers based on the modified  $\mathcal{C}_{RA}(6,L)$ code chain with (a)  $L=50$ and (b) $L=100$. Similar BERs are measured in chains 1 and 2, significantly improving on the BER of chain 3. Also, the BER of chain 3 coincides with the BER of a system using the modified $\mathcal{C}_{RA}(6,L)$ code chain (again demonstrating the UEP property of the CC structure). Similar conclusions can be drawn from Fig. \ref{BERARJA}, where we use a CC structure with $\J=3$ layers based on a modified $\mathcal{C}_{\text{ARJA}}(L)$ code chain  with $L=50$, where connections are placed according to Fig. \ref{CCARJA2}. 

\section{Feasibility of CC  transmission}\label{feasibility}

Without  loss of generality, in this section we consider   CC transmission  using the CC structure based on  $\mathcal{C}(3,6,L)$ code chains in Fig. \ref{CC1} to show that, compared to transmission of independent chain codewords,  CC transmission using connected SC-LDPC code chains only requires some additional memory  in the encoding stage and a different transmission order for the encoded bits. 

\subsubsection{Encoding}

For a given code belonging to the $\mathcal{C}(3,6,L)$ ensemble,  the encoding process can be implemented sequentially using the syndrome former encoder proposed in \cite{pjs+08}. Let $\mathbf{u}^{(i)}$, $i=1,2,\ldots,L$, be a sub-block of $2N\rate(L)$  information bits and $\mathbf{v}^{(i)}$ the corresponding sub-block of $2N$ encoded bits. Using the syndrome former encoder, to compute $\mathbf{v}^{(i)}$ we only need $\mathbf{u}^{(i)}$ and the previously encoded blocks $\mathbf{v}^{(i-1)},\mathbf{v}^{(i-2)},\ldots,$  corresponding to the positions in the SC-LDPC code chain whose variable nodes are connected to the encoded bits $\mathbf{v}^{(i)}$  by the parity check nodes at position $i$.  For example,  $\mathbf{v}^{(1)}$ can be obtained directly from $\mathbf{u}^{(1)}$, $\mathbf{v}^{(2)}$ can be computed given $\mathbf{v}^{(1)}$ and $\mathbf{u}^{(2)}$, and then, for $i=3,4,\ldots,L$, we can compute  $\mathbf{v}^{(i)}$ from  $\mathbf{u}^{(i)}$, $\mathbf{v}^{(i-1)}$,  and $\mathbf{v}^{(i-2)}$. 

Using the syndrome former encoder, the process of encoding $N$ consecutive layers of the CC structure in Fig. \ref{CC1} is essentially equivalent in complexity to encoding $N$ consecutive but independent codewords of the single chain ensemble. The only difference is that, after the first chain  in Fig. \ref{CC1} has been encoded, the encoding of the first sub-block in the $\j$-th chain, $\mathbf{v}^{(1)}_{j}$, not only requires $\mathbf{u}^{(1)}_{j}$, but also $\mathbf{v}_{j-1}^{(\floor{L/2})}$ and $ \mathbf{v}_{j-1}^{(\floor{L/2}+1)}$, $j=2,3,\ldots,\J$, where the subscript $j$ refers to the layer of the CC structure. Similarly, to compute the last encoded sub-block in the $j$-th chain, $\mathbf{v}^{(L)}_{j}$, we also need $\mathbf{v}^{(\floor{L/2}+2)}_{j-1}$ and $\mathbf{v}^{(\floor{L/2}+3)}_{j-1}$. Therefore, compared to encoding $N$ independent chain codewords, we need only  some additional  memory  to store the encoded sub-blocks $\mathbf{v}_{j}^{(\floor{L/2})},\mathbf{v}_{j}^{(\floor{L/2}+1)},\mathbf{v}_{j}^{(\ceil{L/2+2})}$, and $\mathbf{v}_{j}^{(\ceil{L/2+3})}$ that are necessary to encode the chain at layer $j+1$. Note that this additional memory can be reused once each chain is encoded.
\vspace{2mm}

\subsubsection{Window decoding and transmission order}

Efficient  decoding of long SC-LDPC code chains with low decoding delay is based on windowed BP decoding \cite{lscz10,iyengar11}. In a nutshell, decoding is restricted to a window of $W$ positions that `slides' over the graph, exploiting the convolutional structure of the SC-LDPC code parity check matrix: as bits in the left most positions of the window are decoded, the window is shifted  right and new bits are included in the decoding window (see \cite{lscz10} and \cite{iyengar11} for further details). For  sufficiently large $W$, e.g., a window of length $W=10$ positions  for the standard $\mathcal{C}(3,6,L)$  SC-LDPC code chain described in Section \ref{singlechain}, the performance is indistinguishable from a  standard BP decoder, while the delay is much less, since  decoding can be initiated before receiving the entire codeword. 

The same decoding principle can be simply adapted to perform efficient decoding of CC  transmission of SC-LDPC codes. For example, for the CC structure in Fig. \ref{CC1}, the windowed decoder can be initiated at the first position of  chain 1. The window will shift until it reaches the middle positions of the chain, whose bits are better protected since they are also connected to check nodes at the boundary positions of  chain  2. Therefore, to efficiently continue  window decoding of chain 1, channel information from the variable nodes at positions $1$, $2$, and $L$  of  chain  2 must be available.  Note that we need this information even before receiving channel information from  the variable nodes at the remaining positions of  chain  1. Once  chain  1 has been decoded, window decoding of  chain  2 can be started using the information already available at the variable node positions at its boundaries. 

Therefore implementing efficient window decoding for CC transmission reduces to a change in the order in which the encoded bits are transmitted, so that the receiver can have the necessary information at the appropriate time. Clearly, it is necessary for both the transmitter and the receiver to be aware of the transmission order. Returning to the CC structure in Fig. \ref{CC1}, for $\J=2$ layers it would be sufficient to transmit the encoded blocks $\mathbf{v}^{(i)}_{j}$ for $j=1,2$ and $i=1,\ldots,L$ in the following order:
\begin{align}
&\mathbf{v}^{(1)}_{1} \rightarrow  \mathbf{v}^{(2)}_{1} \rightarrow \ldots \rightarrow \mathbf{v}^{(\floor{L/2}+3)}_{1}  \rightarrow
\mathbf{v}^{(1)}_{2}  \rightarrow  \mathbf{v}^{(2)}_{2} \rightarrow  \mathbf{v}^{(L)}_{2} \nonumber \\\nonumber\\
&\rightarrow \mathbf{v}^{(\floor{L/2}+4)}_{1} \rightarrow \ldots  \rightarrow \mathbf{v}^{(L)}_{1} \rightarrow \mathbf{v}^{(3)}_{2} \rightarrow \nonumber \ldots \rightarrow \mathbf{v}^{(L-1)}_{2}.
\end{align}
If we add one layer ($\J=3$), then we would use the same transmission policy between the boundary positions  in chain 3, i.e., $\mathbf{v}^{(1)}_{3}$, $\mathbf{v}^{(2)}_{3}$, and  $\mathbf{v}^{(L)}_{3}$, and the positions $\mathbf{v}^{(\floor{L/2}+3)}_{2}$ and $\mathbf{v}^{(\floor{L/2}+4)}_{2}$ in chain 2.

\section{Conclusions}\label{future}

In this paper, we have reexamined existing results on the analysis of the finite-length performance of SC-LDPC code chains. An important conclusion is that there exists a significant performance improvement as we consider shorter chains that cannot be explained using asymptotic arguments, since the SC-LDPC code threshold quickly saturates with chain length. For short chains, intermediate positions directly benefit from the low-rate terminations at the ends of the graph and this results in finite-length scaling behavior that resembles the single critical point behavior that we find in uncoupled  LDPC block code ensembles.

Based on this result,  a novel transmission scheme (CC transmission) designed to boost the performance of a system using long SC-LDPC code chains was introduced. Using a peeling decoder analysis for the BEC, we have shown that, by connecting consecutive SC-LDPC  code chains rather than transmitting the codewords corresponding to each one independently, we obtain a significant performance improvement  with only a minor change in the order in which coded bits are transmitted and some additional memory requirements at the encoder. The design of the CC structure relies on the creation of robust regions in the middle of each SC-LDPC code chain and, to this end, strategies have been presented for different code ensembles, each with different trade-offs between design rate and the fraction of code chains for which improved performance is achieved.  CC transmission is illustrated using several representative protograph-based SC-LDPC code ensembles proposed to date in the literature, and computer simulation results verifying the claimed performance improvements are presented. An interesting open research problem is to combine CC transmission with recently proposed system design approaches that also aim to enhance the use of SC-LDPC codes in modern communication system applications \cite{Alex15,Aref15}.


\bibliography{allbib20_03.bib}
\bibliographystyle{IEEEtran}

\end{document}